\documentclass[twocolumn,prb,showpacs,floatfix]{revtex4}

\usepackage{amssymb}
\usepackage{amsmath}
\usepackage{epsfig}
\usepackage{graphics}
\usepackage[dvips]{color}

\begin{document}

\title{Cyclotron motion in graphene}

\author{John Schliemann}

\affiliation{Institute for Theoretical Physics, University of 
Regensburg, D-93040 Regensburg, Germany}

\date{\today}

\begin{abstract}
We investigate cyclotron motion in graphene monolayers considering both
the full quantum dynamics and its semiclassical limit reached at high
carrier energies. Effects of {\em zitterbewegung} due to the two
dispersion branches of the spectrum dominate the 
irregular quantum motion at low energies
and are obtained as a systematic correction to the semiclassical case.
Recent experiments are shown to operate in the semiclassical regime.
\end{abstract}

\maketitle

\section{Introduction}

The experimental discovery of graphene, i.e. single layers of graphite,
has rapidly led to an extraordinary vast and still growing interest in this
material, both experimental and theoretical
\cite{Novoselov04,Geim07,CastroNeto07}. Compared to conventional
two-dimensional electronic systems, the peculiar properties 
of graphene mainly stem from its linear dispersion near the Fermi energy,
and the chiral nature of electronic states entangling the momentum and
sublattice degree of freedom \cite{Geim07,CastroNeto07}.
In particular, studies of carrier transport in perpendicular magnetic fields
have revealed unusual features like a cyclotron mass being
proportional to the square root of the particle density 
\cite{Novoselov05}, and, most spectacular, a quantum Hall effect
occuring at half-integer filling factors
\cite{Novoselov05,Zhang05,Novoselov07}. Another partially related
feature of graphene is its spectrum of Landau levels being
non-equidistant in energy and proportional to the square root
of the magnetic field,
\cite{Novoselov05,Zhang05,Novoselov07,Deacon07,Jiang07a,Jiang07b},
properties also not shared by usual two-dimensional electron gases in
semiconductor structures.

In this communication we analyze
the cyclotron motion in graphene considering both the full quantum
dynamics as well as its semiclassical limit. The latter findings can directly
be compared with the the results of Shubnikov-de Haas measurements
reported on in Ref.~\cite{Novoselov05}. We also comment on a
recent preprint by Rusin and Zawadzki
studying cyclotron motion in graphene
which appeared while the present work was reaching completion\cite{Rusin07a}. 
These authors concentrate on the full quantum dynamics using a numerical 
approach analogous to very recent work \cite{Schliemann07a}
on two-dimensional electron gases in semiconductors,
and  they stress the role of {\em zitterbewegung}
\cite{Schliemann05,Schliemann06,Zawadzki05,Nikolic05,Shen05,Cserti06,Brusheim06,Rusin07b,Trauzettel07,Rusin07c,Schliemann07b,Winkler07,Bernardes07,Zulicke07a}. 
As we shall see below, signatures of 
{\em zitterbewegung} can also be seen in semiclassical corrections 
to the classical limit of the underlying quantum dynamics.

In the following section \ref{quantum} we investigate 
the full quantum dynamics of the system using both 
a numerical and an anyltical approach. The latter one is based on the
analogy of the Hamiltonian to the Jaynes-Cummings model steming from
quantum optics. This analogy is the starting point for the analysis
of the semiclassical limit to be discussed in section \ref{semicl}.
There we also compare our results with the 
Shubnikov-de Haas expriments by
Novoselov {\em et al.} \cite{Novoselov05}. 
Section \ref{concl} contains our conclusions.

\section{Quantum Dynamics}
\label{quantum}

For a graphene sheet in a perpendicular magnetic field, the single-particle 
states around one of one of the two inequivalent corners of the first
Brillouin zone are described by \cite{CastroNeto07}
\begin{equation}
{\cal H}=v\left(\tau^{z}\pi_{x}\sigma^{x}+\pi_{y}\sigma^{y}\right)
\label{defhamil}
\end{equation}
with (using standard notation) $\vec\pi=\vec p+e\vec A/c$ and 
$v\approx 10^{6}{\rm m/s}$. The Pauli matrices describe the sublattice 
or pseudospin degree
of freedom, and the Zeeman coupling to the physical electron spin has been 
neglected. The label $\tau^{z}=\pm 1$ determines which corner of the Brillouin
zone in considered; in what folows we shall concentrate on 
 $\tau^{z}=1$ The Heisenberg equations of motion read
\begin{eqnarray}
\frac{d}{dt}\vec\pi_{H}(t) & = & \frac{\hbar v}{\ell^{2}}\vec\sigma_{H}(t)
\times\vec e_{z}
\label{eom1}\\
\frac{d}{dt}\vec\sigma_{H}(t)& = & \frac{2v}{\hbar}\vec\pi_{H}(t)
\times\vec\sigma_{H}(t)
\label{eom2}
\end{eqnarray}
where $\ell=\sqrt{\hbar c/|eB|}$, $(-e)B>0$, is the the magnetic length,
and $\vec e_{z}$ is the unit vector along the $z$-direction. The magnetic field
$\vec B=\nabla\times\vec A=B\vec e_{z}$ is assumed to point along the negative 
$z$-direction, $B<0$. The position operator $\vec r=(x,y)$ 
can be given in terms of
the kinetic momentrum $\pi$ via the usual relations
\begin{eqnarray}
x & = & x_{0}+\frac{c}{eB}\pi_{y}\,,\\
y & = & y_{0}-\frac{c}{eB}\pi_{x}\,,
\end{eqnarray}
where $\vec r_{0}=(x_{0},y_{0})$ is conserved, 
$[{\cal H},\vec r_{0}]=0$. For cyclotron motion in a
usual non-interacting two-dimensional electron gas, the vector
$\vec r_{0}$ describes the center of the classical circular orbits.
As we shall see below, this is also the case for the classical limit
of cyclotron motion in graphene.

Defining the usual bosonic operators
\begin{equation}
a=\frac{1}{\sqrt{2}}\frac{\ell}{\hbar}\left(\pi_{x}+i\pi_{y}\right)
\quad,\quad a^{+}=(a)^{+}
\end{equation}
fulfilling $[a,a^{+}]=1$ the Hamiltonian reads
\begin{equation}
{\cal H}=\frac{\hbar v}{\ell}\sqrt{2}\left(a\sigma^{-}+a^{+}\sigma^{+}\right)
\label{hamil}
\end{equation}
where $\sigma^{\pm}=(\sigma^{x}\pm i\sigma^{y})/2$. The energy scale of
this Hamiltonian is given as a function of the magnetic field by
\begin{equation}
\frac{\hbar v}{\ell}\approx 26 {\rm meV}\sqrt{\frac{B}{\rm Tesla}}\,,
\end{equation}
while its length scale is the usual magnetic length,
$\ell=257 \AA / \sqrt{B/ {\rm Tesla}}$. The well-known eigenstates
\cite{CastroNeto07} of the Hamiltonian (\ref{hamil}) are given by 
$|0,\uparrow\rangle$ with energy $\varepsilon_{0}=0$ and, for $n>0$,
\begin{equation}
|n,\pm\rangle=\frac{1}{\sqrt{2}}\left(
|n,\uparrow\rangle\pm|n-1,\downarrow\rangle\right)
\label{eigen}
\end{equation}
with energy $\varepsilon_{n}^{\pm}=\pm(\hbar v/\ell)\sqrt{2n}$.
Here $n$ is the Landau level index, and the arrows are obvious standard 
notation for the sublattice spin states.

\subsection{Numerical Approach}

The operator-valued equations of motion (\ref{eom1}), (\ref{eom2})
do not seem to allow for a full exlicit solution.
However, it is straightforward though somewhat tedious to numerically
evaluate the time evolution of momentum, position, and spin operators.
A similar approach was performed recently in Ref.~\cite{Schliemann07a}
investigating cyclotron motion in semiconductor quantum wells with
spin-orbit coupling. In fact, the present case of graphene is technically
clearly simpler than the previous one and was also studied very recently in
Ref.~\cite{Rusin07a}. For such numerical simulation it is convenient to
work in the Landau gauge $\vec A=(0,Bx,0)$ with the initial
state $|\psi\rangle$ being a direct product of an orbital and a spin state,
\begin{equation}
|\psi\rangle=|\phi\rangle\left(
\begin{array}{c}
\kappa \\
\lambda
\end{array}
\right)\,,
\label{init}
\end{equation}
where the spinor components are related to the usual polar angles
$\vartheta$, $\varphi$ of the initial pseudospin direction
via $\kappa=\exp(-i\varphi/2)\cos(\vartheta/2)$, $\lambda=\exp(i\varphi/2)\sin(\vartheta/2)$. As a generic initial 
orbital state we consider
\begin{equation}
\langle\vec r|\phi\rangle=\frac{1}{\sqrt{\pi}d}e^{-\frac{r^{2}}{2d^{2}}+ik_{0}y}\,,
\label{initorb}
\end{equation}
i.e. a normalized Gaussian wave packet of spatial width $d$ and initial
momentum $\hbar k_{0}$ along the $y$-axis, i.e. the direction of translational 
invariance of the Hamiltonian. The initial position of the particle is at
the origin, $\langle\psi|\vec r|\psi\rangle=0$, and its energy is given by
\begin{equation}
E=\langle\psi|{\cal H}|\psi\rangle=\hbar vk_{0}\frac{1}{i}
\left(\bar\kappa\lambda-\kappa\bar\lambda\right)
\end{equation} 
with a quantum mechanical uncertainty of
\begin{eqnarray}
\left(\Delta{\cal H}\right)^{2} & = & \langle\psi|{\cal H}^{2}|\psi\rangle
-\langle\psi|{\cal H}|\psi\rangle^{2}\nonumber\\
 & = & \left(\frac{\hbar v}{\ell}\right)^{2}
\Biggl(\frac{\ell^{2}}{d^{2}}+\frac{d^{2}}{2\ell^{2}}
-\left(|\kappa|^{2}-|\lambda|^{2}\right)\nonumber\\
 & & \qquad\quad\quad+k_{0}^{2}\ell^{2}
\left(1+\left(\bar\kappa\lambda-\kappa\bar\lambda\right)^{2}\right)\Biggr)
\end{eqnarray}
Note that the initial state (\ref{init}) has in general non-vanishing
overlap with single-particle eigenstates of the form (\ref{eigen})
of both positive and negative energy. As to be discussed below, this
fact leads to additional oscillations in the time evolution
that can be viewed as {\em zitterbewegung}
\cite{Rusin07a,Schliemann05,Schliemann06}.

We emphasize that,
by the very construction of the Hamiltonian (\ref{defhamil}),
the wave number $k_{0}$ is to be interpreted relatively to the
wave vector of the chosen corner of the first Brillouin zone
\cite{CastroNeto07}. 
The authors of Ref.~\cite{Rusin07a}, however, consider a time
evolution of an initial state of the form (\ref{init}) under
the simultaneous action of the Hamiltonians of both inequivalent
corners of the Brillouin zone, a theoretical modeling whose physical 
meaning remains rather unclear. 
Moreover, these two Hamiltonians are assumed to differ
just in a global sign, i.e. one Hamiltonian is the negative of the other,
which is at odds with the microscopic tight-binding description of
graphene \cite{CastroNeto07}. 

\begin{center}
\begin{figure}
\epsfig{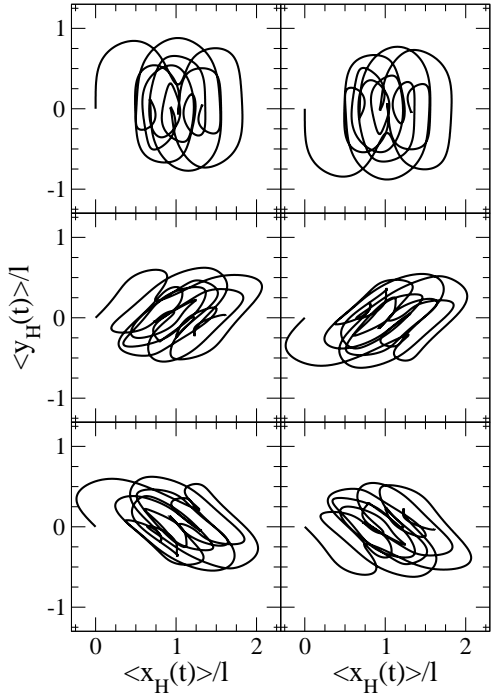}
\caption{Orbital dynamics of a wave packet of initial width $d=1.0\ell$ and
group wave number $k_{0}=1.0/ \ell$. In all cases the sublattice spin
lies initially in the $xy$-plane, i.e. $\vartheta=\pi/2$ leading to
$|\kappa|=|\lambda|$, and the total simulation time is always
$50\ell/v$. In the left and right top panel, the sublattice spin
is initially collinear with the momentum with $\varphi=\pi/2$ and 
$\varphi=3\pi/2$,
respectively. In the middle panels we have $\varphi=\pi/4$ (left)
and  $\varphi=5\pi/4$ (right) as initial conditions, while in the
bottom panels $\varphi=3\pi/4$ and $\varphi=7\pi/4$ were used}
\label{fig1}
\end{figure}
\end{center}

For an infinite sheet of graphene it is natural to consider initial conditions
where both sublattices have the same quantum mechanical weight,
$|\kappa|=|\lambda|$, i.e. the sublattice or pseudospin lies initially
in its $xy$-plane. Fig.~\ref{fig1} shows several examples
for numerically evaluated trajectories 
$\langle\vec r_{H}(t)\rangle:=\langle\psi|\vec r_{H}(t)|\psi\rangle$
with this type of initial condition. For all further details
of these conceptually straightforward but technically somewhat involved
numerical simulations we refer to Refs.~\cite{Rusin07a,Schliemann07a}.
In the two top panels of Fig.~\ref{fig1} the sublattice spin is
initially collinear with the momentum. These simulations can be compared with 
Fig.~5 of Ref.~\cite{Rusin07a} where the authors consider a
time evolution under a {\em single} Hamiltonian (not two of them)
given by Eq.~(\ref{defhamil}). Indeed, all results given in that figure
are essentially reproduced by our own simulations underlying the present work.

The remaining panels of Fig.~\ref{fig1} contain simulations where the
sublattice spin is in the initial state not collinear with the momentum.
Note that the initial velocity 
$\langle\vec v_{H}\rangle=\langle\dot\vec r_{H}\rangle$
is determined by the initial direction of the sublattice spin via
\begin{equation}
\vec v=\frac{i}{\hbar}[{\cal H},\vec r]=v\vec\sigma\,.
\label{velocity}
\end{equation}
Moreover, further simulations show that there is no dynamics
in the initial direction of the momentum, i.e. 
$\langle y_{H}(t)\rangle=0$, if the sublattice spin is
initially perpendicullar to it, i.e. in the $xz$-plane. In such a case
also no sublattice spin component collinear with the
initial momentum develops in the time evolution, 
$\langle\sigma^{y}_{H}(t)\rangle=0$. These observations were partially
already made in Ref.~\cite{Rusin07a} and can be understood 
from the equations of motion (\ref{eom1}), (\ref{eom2}).
In general, the trajectories seen in Fig.~\ref{fig1} are rather
irregular which is a consequence of the many different excitation
frequencies being present in the spectrum at low energies.
In Ref.~\cite{Rusin07a} contributions to the time evolution
involving transition frquencies between states of positive and negative
energy have been regarded as an effect of {\em zitterbewegung}.
Indeed, as to be shown below, such type of {\em zitterbewegung} also occurs
as a correction to the classical limit.

As already mentioned in Refs.~\cite{Rusin07a, Schliemann07a}
the numerical approach discussed above is technically limited to initial states
having significant overlap with rather low Landau levels only, i.e. the
method is restricted to the regime dominated by quantum effects.
In the following we shall therefore explore the semiclassical limit
using an analytical approach.

\subsection{Analogy to the Jaynes-Cummings model}

Further analytical progress regarding the full quantum dynamics
can be made by exploiting the fact that the
Hamiltonian (\ref{hamil}) is formally equivalent to the Jaynes-Cummings model 
for atomic transitions in a radiation field. A similar observation
has been made recently in Ref.~\cite{Winkler07} for the two-dimensional
electron gas in semiconductor quantum wells with Rashba spin-orbit coupling.
The Jaynes-Cummings model has been studied very
intensively in theoretical quantum optics, and the time evolution of
orbital and spin operators has been obtained in terms of analytical but
rather implicit expressions \cite{Ackerhalt75,Barnett97}.
Using the method described in Refs.~\cite{Ackerhalt75,Barnett97} one can solve
for the time-dependent position operators in the Heisenberg picture as
\begin{eqnarray}
 & & x_{H}(t)+iy_{H}(t)=x_{0}+iy_{0}\nonumber\\
 & & \qquad+\frac{i\ell e^{-i\omega_{+}t}}{\omega_{-}-\omega_{+}}
\left(\omega_{-}\left(\kappa_{x}+i\kappa_{y}\right)\ell
-2\frac{v}{\ell}\sigma^{+}\right)
\nonumber\\
 & & \qquad-\frac{i\ell e^{-i\omega_{-}t}}{\omega_{-}-\omega_{+}}
\left(\omega_{+}\left(\kappa_{x}+i\kappa_{y}\right)\ell
-2\frac{v}{\ell}\sigma^{+}\right)\,,
\label{acker}
\end{eqnarray}
with $\vec\kappa=\vec\pi/ \hbar$, and the operator-valued frequencies 
$\omega_{\pm}$ are given by
\begin{equation}
\hbar\omega_{\pm}=-{\cal H}\pm\sqrt{{\cal H}^{2}
+2\left(\frac{\hbar v}{\ell}\right)^{2}}\,.
\end{equation}
All operators on the r.h.s of Eq.~(\ref{acker}) are in the Schr\"odinger
ppicture, i.e. at time $t=0$.
As in the case of the two-dimensional electron gas with Rashba spin-orbit
coupling investigated previously \cite{Winkler07,Schliemann07a}, the result
(\ref{acker}) is still rather implicit and difficult to evaluate
for a given initial state, mainly due to the operator character of
the frequencies $\omega_{\pm}$. However, as we shall see in next section,
the above result provides a very natural access to the classical limit
of the dynamics including semiclassical corrections.

\section{Semiclassical Limit}
\label{semicl}

Cyclotron motion of massful electrons is in the simplest case just
described ba a kinetic (effective-)mass term, and the quantum mechanical
and the classical equations of motion coincide, the latter ones being the 
obvious and well-defined classical limit of the former. However, the classical
limit of the Hamiltonian (\ref{defhamil}) describing massless fermions
appears {\em prima vista} not as obvious. 
In general, the classical limit of a quantum system is approached
in the limit of high energies. For our problem here this means that
the energy $E=\langle{\cal H}\rangle$ of electron must be large compared
to the charcteristic energy scale $\hbar v/ \ell$ of the Hamiltonian, which    
is equivalent to the condition
$\langle\vec\kappa\cdot\vec\sigma\rangle\ell\gg 1$. In what follows
we shall assume an initial state of the form (\ref{init}) with
momentum and sublattice spin being parallel to each other.
This leads to the condition
\begin{equation}
k_{0}\ell\gg 1\,.
\label{cond1}
\end{equation}
This very natural classical limit of the graphene model
in a perpendicular magnetic field corresponds to the 
``strong-coupling scenario'' discussed in Ref.~\cite{Zulicke07a}.
We note that a similar limit is reached for negative energies of large
modulus. Here momentum and sublattice spin are initially antiparallel. 

In the limit of large energies $E$, the operator-valued
frequencies $\omega_{\pm}$ can be replaced with classical variables,
\begin{equation}
\hbar\omega_{\pm}\mapsto-E\pm\sqrt{E^{2}
+2\left(\frac{\hbar v}{\ell}\right)^{2}}\,.
\end{equation}
Then, treating also the the quantites $\vec r$, $\vec r_{0}$ as well as
$\vec\kappa$, $\vec\sigma$ as classical 
variables (not as operators) one derives from the full quantum result
(\ref{acker}) in the above classical limit
\begin{eqnarray}
 & & x(t)-x_{0}=\frac{\ell}{\omega_{-}-\omega_{+}}\nonumber\\
 & & \qquad\cdot\Biggl[\omega_{-}\left(\kappa_{x}\ell\sin(\omega_{+}t)
-\kappa_{y}\ell\cos(\omega_{+}t)\right)\nonumber\\
 & & \qquad+\frac{v}{\ell}\left(-\sigma^{x}\sin(\omega_{+}t)
+\sigma^{y}\cos(\omega_{+}t)\right)\nonumber\\
 & & \qquad-\omega_{+}\left(\kappa_{x}\ell\sin(\omega_{-}t)
-\kappa_{y}\ell\cos(\omega_{-}t)\right)\nonumber\\
 & & \qquad-\frac{v}{\ell}\left(-\sigma^{x}\sin(\omega_{-}t)
+\sigma^{y}\cos(\omega_{-}t)\right)\Biggr]\,,\\
 & & y(t)-y_{0}=\frac{\ell}{\omega_{-}-\omega_{+}}\nonumber\\
 & & \qquad\cdot\Biggl[\omega_{-}\left(\kappa_{x}\ell\cos(\omega_{+}t)
+\kappa_{y}\ell\sin(\omega_{+}t)\right)\nonumber\\
 & & \qquad-\frac{v}{\ell}\left(\sigma^{x}\cos(\omega_{+}t)
+\sigma^{y}\sin(\omega_{+}t)\right)\nonumber\\
 & & \qquad-\omega_{+}\left(\kappa_{x}\ell\cos(\omega_{-}t)
+\kappa_{y}\ell\sin(\omega_{-}t)\right)\nonumber\\
 & & \qquad+\frac{v}{\ell}\left(\sigma^{x}\cos(\omega_{-}t)
+\sigma^{y}\sin(\omega_{-}t)\right)\Biggr]\,.
\end{eqnarray}
Again the quantities $\vec\kappa$, $\vec\sigma$ in the 
rectangular brackets on the r.h.s. are at time $t=0$. 
Moreover, in the classical limit $E\gg\hbar v/ \ell$ 
we have $\hbar\omega_{-}\approx-2E$ and 
$\hbar\omega_{+}\approx(\hbar v/ \ell)^{2}/E$. It is instructive to rewrite
the latter expression in the usual form of a cyclotron frequency 
$\hbar\omega_{+}=:\omega_{c}=|eB|/m_{c}c$. Here the cyclotron mass
is given by the well-known semiclassical expression
\cite{Ashcroft76,Novoselov05}
$m_{c}=(\hbar^{2}/2\pi)\partial S/ \partial E=E/v^{2}$ where
$S=\pi k_{0}^{2}$ is the area enclosed by a cyclotron orbit, and
$E=\hbar vk_{0}$. Now, expanding the above expressions in first
order in $(\hbar v/ \ell)/E$ one finds 
\begin{eqnarray}
x(t)-x_{0} & = & \kappa_{x}\ell^{2}\sin(\omega_{c}t)
-\kappa_{y}\ell^{2}\cos(\omega_{c}t)\nonumber\\
 & + & \frac{\hbar v}{E}
\Bigl(-\sigma^{x}\sin(\omega_{c}t)+\sigma^{y}\cos(\omega_{c}t) \nonumber\\
 & & -\sigma^{x}\sin(2Et/\hbar)-\sigma^{y}\cos(2Et/\hbar)\Bigr)\,,
\label{semix}\\
y(t)-y_{0} & = & \kappa_{x}\ell^{2}\cos(\omega_{c}t)
+\kappa_{y}\ell^{2}\sin(\omega_{c}t)\nonumber\\
 & - & \frac{\hbar v}{E}
\Bigl(\sigma^{x}\cos(\omega_{c}t)+\sigma^{y}\sin(\omega_{c}t) \nonumber\\
 & & -\sigma^{x}\cos(2Et/\hbar)+\sigma^{y}\sin(2Et/\hbar)\Bigr)\,.
\label{semiy}
\end{eqnarray}
The first lines of the above r.h.s. describe the classical cyclotron motion
\cite{Ashcroft76,Schliemann07a} with an energy-dependent cycloron mass
\cite{Novoselov05}. The other contribution are lowest-order
semiclassical corrections to this classical limit. In particular, the
terms in the last lines oscillate with the large frequency
$2E/ \hbar\gg\omega_{c}$ which equals the energy separation 
$2E=2\hbar vk_{0}$ between states of positive and negative energy
at given wave vector in the absence of a magnetic field \cite{CastroNeto07}.
Therefore, these semiclassical correction can be viewed as an effect
of {\em zitterbewegung} \cite{Rusin07a,Schliemann05,Schliemann06}.
We note that in connection with graphene thne term {\em ``zitterbewegung''}
was also used recently to express the fact that the velocity operator
(\ref{velocity}) fails to commute with the Hamiltonian (\ref{defhamil})
\cite{Katsnelson06}.

So far we have concentrated on semiclassical dynamics at large positive 
energies $E\gg\hbar v/\ell$. In the analogous case of negative energies of
large modulus one obtains a similar result with the frequencies 
$\omega_{\pm}$ being interchanged, $\omega_{+}\approx2|E|$, 
$\omega_{-}\approx-\omega_{c}$.

Finally we note the equivalence of the following three conditions:
\begin{eqnarray}
 & E\gg\hbar\omega_{c}=\frac{\left(\frac{\hbar v}{\ell}\right)^{2}}{E} &
\label{cond2}\\
\Leftrightarrow & E^{2}=(\hbar vk_{0})^{2}\gg
\left(\frac{\hbar v}{\ell}\right)^{2} & \label{cond3}\\
\Leftrightarrow & (k_{0}\ell)^{2}\gg 1& \label{cond4}\,.
\end{eqnarray}
Thus, the condition (\ref{cond2}) is fulfilled whenever
(\ref{cond1}) is valid. On the other hand, the condition (\ref{cond1})
can be rewritten as $r_{c}\gg\ell$ with $r_{c}=k_{0}\ell^{2}$ being the
classical cyclotron radius. This is indeed the usual textbook criterion 
for the validity of semiclassical approximations to cyclotron dynamics
in solids \cite{Ashcroft76}
and confirms again our
above strategy of obtaining the semiclassical limit.

Let us now compare our results with the Shubnikov-de Haas measurements
by Novoselov {\em et al.} \cite{Novoselov05}. The data presented in
Fig.~2a of Ref.~\cite{Novoselov05} was obtained at an electron density of
$n\approx 4.3\cdot 10^{12}{\rm cm^{2}}$ and magnetic fields of up to about
$B=10{\rm T}$, corresponding to a Fermi wave vector of
$k_{f}=\sqrt{\pi n}\approx 3.7\cdot10^{6}{\rm cm^{-1}}$ and a magnetic
length of $\ell\gtrsim81\AA$. Thus, $k_{f}\ell\gtrsim3$, and the criterion
(\ref{cond1}) is still reasonably fulfilled even for
the highest fields used in those measurements. Correspondingly,
we have $(\hbar v/ \ell)/E\lesssim 0.33$ with $E=\hbar vk_{f}$, showing
that the semiclassical correction obtained in Eqs.~(\ref{semix}),(\ref{semiy})
are sufficiently suppressed which allows for an interpretation of the 
experimental data in purely classical terms. At higher magnetic fields,
quantum effects dominate, and quantized Hall transport is observed
\cite{Novoselov05,Zhang05,Novoselov07}.

\section{Conclusions}
\label{concl}

We have investigated cyclotron motion in graphene monolayers considering both
the full quantum dynamics and its semiclassical limit obtained for high
carrier energies. At low energies, the quantum dynamics leads to rather
irregular particle trajectories dominated by  
effects of {\em zitterbewegung} due to the two
dispersion branches of the spectrum.
The semiclassical limit of the system is obtained using the
analogy of the Hamiltonian with the Jaynes-Cummings model
\cite{Winkler07,Schliemann07a}. A similar analysis of the classical limit
can be done for cyclotron motion in semiconductor quantum wells with
spin-orbit interaction \cite{Schliemann07a,Winkler07}.
The semiclassical limit of cyclotron dynamics in graphene
is described by the usual
cyclotron frequency and the characteristic frequency of 
{\em zitterbewegung} which occurs as a semiclassical correction.
Recent experiments are shown to operate in the semiclassical regime.

\acknowledgments{I thank M.~I. Katsnelson, M. Trushin, and 
U. Z\"ulicke for useful discussions. 
This work was supported by DFG via SFB 689  
``Spin Phenomena in reduced Dimensions''.}

\end{document}